# Current-driven resonant excitation of magnetic vortex

Shinya Kasai, [1] Yoshinobu Nakatani, [2] Kensuke Kobayashi, [1] Hiroshi Kohno, [3] Teruo Ono [1]

[1]*Institute for Chemical Research, Kyoto University, Uji 611-0011, Japan*

[2]*University of Electro-communications, Chofu 182-8585, Japan*

[3]*Graduate School of Engineering Science, Osaka University, Toyonaka 560-8531, Japan*

**Abstract**

A magnetic vortex core in a ferromagnetic circular nanodot has a resonance frequency originating from the confinement of the vortex core. By the micromagnetic simulation including the spin-transfer torque, we show that the vortex core can be resonantly excited by an AC (spin-polarized) current through the dot and that the resonance frequency can be tuned by the dot shape. The resistance measurement under the AC current successfully detects the resonance at the frequency consistent with the simulation.

The manipulation of magnetization by spin currents is a key technology for future spintronics [1-14]. The underlying physics is that spin currents can apply a torque on the magnetic moment when the spin direction of the conduction electrons has a relative angle to the local magnetic moment. This leads us to the hypothesis that any type of spin structure with spatial variation can be excited by a spin-polarized current in a ferromagnet. The ideal example of such a noncollinear spin structure is a curling magnetic structure ("magnetic vortex") realized in a ferromagnetic circular nanodot. Although this structure was theoretically predicted long ago [15], it was only recently confirmed by microscopic experiments that such a vortex exists with a nanometer-scale core where the curling magnetization becomes out-of-plane [16,17]. Subsequent intensive studies have clarified that after switching off an in-plane magnetic field, a vortex core exhibits a spiral precession around the dot center during the relaxation process [18-20]. Thus, the nanodot functions as a resonator for vortex core motion. Thus far, magnetostatic interactions triggered by an external magnetic field have dominated the study of vortex dynamics; however, the abovementioned concept—vortex manipulation by a spin-polarized current—has not been explored.

Here, we demonstrate that a magnetic vortex core can be resonantly excited by an AC current through a ferromagnetic circular dot when the current frequency is tuned to the eigenfrequency originating from the confinement of the vortex core in a dot. Our micromagnetic simulations with the spin-transfer effect reveal in detail the motion during the excitation; an excited vortex core draws a spiral trajectory to settle in a steady orbital around the dot center. We succeeded in detecting the predicted resonance by resistance measurements. We observed efficient excitation by an electric current due to the resonant nature and tunability of the resonance frequency based on the dot shape. By micromagnetic simulations including the spin-transfer effect, we show below that an AC spin-polarized current with the eigenfrequency of the resonator can resonantly excite a magnetic vortex core. Then, we present the results of the experimental detection of the resonance of a vortex core, as predicted by the simulation.

Figure 1 shows a scanning electron microscope image of the sample and the schematic configuration used for the measurements. The samples were fabricated on thermally oxidized Si substrates by the lift-off method in combination with e-beam lithography. Each sample consists of a Permalloy ($Fe_{19}Ni_{81}$) dot and two 50 nm-thick Au wide electrodes. The thickness of the dot $h$ is 40 nm, and the radius $r$ is varied to be $r =$ 410, 530, and 700 nm. The existence of a vortex core in each dot was confirmed by conventional magnetic force microscopy. The current-induced dynamics of the vortex core was calculated by the micromagnetic simulations based on the Landau-Lifshits-Gilbert (LLG) equation with a spin-transfer term [21,22]. The modified LLG equation is given by

$$\frac{\partial \boldsymbol{m}}{\partial t} = -\gamma_0 \boldsymbol{m} \times \boldsymbol{H}_{\text{eff}} + \alpha \boldsymbol{m} \times \frac{\partial \boldsymbol{m}}{\partial t} - (\boldsymbol{u}_s \cdot \nabla)\boldsymbol{m} \ , \tag{1}$$

where $\boldsymbol{m}$ is a unit vector along the local magnetization; $\gamma_0$, the gyromagnetic ratio; $\boldsymbol{H}_{\text{eff}}$, the effective magnetic field including the exchange and the demagnetizing fields; and $\alpha$, the Gilbert damping constant. The last term represents the spin-transfer torque, which describes the effect of spin transfer from conduction electrons to localized spins. This spin-transfer effect is a combined effect of the spatial nonuniformity of magnetization and the current flow. The vector $\boldsymbol{u} = -j P g \mu_B/(2eM_s)$, which has the dimension of velocity, is essentially the spin current associated with the electric current in a ferromagnet, where $\boldsymbol{j}$ is the current density; $P$, the spin polarization of the current; $g$, the g-value of an electron; $\mu_B$, the Bohr magneton; $e$, the electronic charge; and $M_s$, the saturation magnetization. In the simulation, an electric current in a dot was assumed to be uniform, and an AC electric current in the form of $j = J_0 \sin 2\pi f t$ was applied, where $J_0$ is the current density; $f$, the frequency of the AC current; and $t$, the time. The dot was divided into rectangular prisms of $4 \times 4 \times 40$ nm$^3$; the magnetization in each of these was assumed to be constant. The typical material parameters for Permalloy were used: $M_s = 1$ T, the exchange stiffness constant $A = 1.0 \times 10^{-11}$ J/m, $P = 0.7$ [23], and $\alpha = 0.01$.

First, we determined the eigenfrequency $f_0$ of the vortex core precession in the dot by calculating the free relaxational motion of the vortex core from the off-centered position. The eigenfrequency depends on the aspect ratio $h/r$ (the height $h$ to the radius $r$) of the dot [18]. Then, the simulations were performed by applying an AC current at a given frequency $f$ in the absence of a magnetic field. Figure 2(a) shows the time evolution of the core position when an AC current ($f = f_0 = 380$ MHz and $J_0 = 3 \times 10^{11}$A/m$^2$) is applied  to a dot with $r = 410$ nm and $h = 40$ nm. Once the AC current is applied, the vortex core first moves in the direction of the electron flow or spin current. This motion originates from the spin-transfer effect. The off-centered core is then subjected to a restoring force toward the dot center. Further, because of the gyroscopic nature of the vortex (the vortex moves perpendicular to the force), the core makes a circular precessional motion around the dot center [18]. The precession is amplified by the current to reach a steady orbital motion where the spin-transfer from the current is balanced with the damping, as depicted in Fig. 2(a). The direction of the precession depends on the direction of the core magnetization as in the motion induced by the magnetic field [18,24]. It should be noted that the radius of the steady orbital on resonance is larger by more than an order of magnitude as compared to the displacement of the vortex core induced by a DC current of the same amplitude [24]. Thus, the core is efficiently excited by the AC current due to resonance. We verified that the resonant excitation of the vortex core presented above was not observed in the micromagnetic simulations without the spin-transfer term even if we included a magnetic field generated by an AC current into the simulation, which indicates that the vortex core is excited not by the Oersted field but by the spin-polarized current.

Figure 2(b) shows the time evolutions of the $x$ position of the vortex core for three different excitation frequencies $f = 250$, 340, and 380 MHz. The steady state appears after around 30 ns on resonance ($f = 380$ MHz). For $f = 340$ MHz slightly off the resonance, the amplitude beats first, and then the steady state with smaller amplitude appears. The vortex core shows only a weak motion for $f = 250$ MHz, which is quite far from the resonance. The displacement amplitude for the nonresonant response is essentially the same as the case of the DC current application [24]. Figure 2(c) shows the radii of the steady orbitals as a function of the current frequency for the dots with $r = 410$, 530, and 700 nm. Each dot exhibits the resonance at the eigenfrequency of the vortex motion.

Our experimental detection method for the resonant excitation of a vortex core is based on the difference in resistance of the dot between the on- and off-resonance states as described below. In general, the resistance of ferromagnetic metals depends on the relative angle between the magnetization and the measuring current—known as the anisotropic magnetoresistance (AMR) effect. Figure 3(a) shows the results of the magnetoresistance measurements at room temperature for the dot with $r = 700$ nm. The resistance was measured by a lock-in technique using a current of 10 µA at 223 Hz. The magnetic field was applied perpendicular to the measuring current ($H \perp I$, the result is indicated by the blue curve) or parallel to the measuring current in the dot plane ($H \mathbin{/\mkern-5mu/} I$, the result is indicated by the red curve). The spin structure of the dot for each state is also indicated in the figure. Figure 3(a) clearly indicates that the resistance of the dot is highly correlated with the core position because of the AMR effect. The key feature is that the resistance change for $H \perp I$ ($|\Delta R_\perp|$) is larger than that for $H \mathbin{/\mkern-5mu/} I$ ($|\Delta R_{/\mkern-5mu/}|$), as seen in the plot of $|\Delta R_{/\mkern-5mu/}| - |\Delta R_\perp|$ as a function of $H$ (Fig. 3(b)). This difference in the resistance change results from the symmetry breaking of the system because of the two electrodes attached to the dot. When the core is on resonance and the measurement time is considerably longer than the period of the core orbital motion, the experimentally measured resistance is the average value of the resistances for all the core positions in the orbital shown in Fig. 2(a). This averaged resistance on resonance is expected to be smaller than that for the off-resonance state, in which the core remains around the dot center because ($|\Delta R_{/\mkern-5mu/}| - |\Delta R_\perp|) < 0$, as shown in Fig. 3(b). We detect the resonance in this manner.

We measured the resistance of the dot while an AC excitation current was passed through it at room temperature in the configuration shown in Fig. 1. The resistance measurements were performed by conventional lock-in techniques using a current of 15 µA with a frequency of 223 Hz. The amplitude of the AC excitation current was $3 \times 10^{11}$ A/m$^2$. Figure 4(a) shows the resistances as a function of the frequency of the AC excitation current for the dots with three different radii $r$ = 410, 530, and 700 nm. A small but clear dip is observed for each dot; this signifies the resonance. Since the observed dip originates from the AMR effect averaged over the vortex orbital, the maximum signal corresponding to the core motion along the dot edge is expected to be about ($|\Delta R_{//}|$ - $|\Delta R_{\perp}|$)/2 $\approx$ –10 m$\Omega$ on the basis of the result for $r$ = 700 nm shown in Fig. 3(b). Thus, the observed signal amplitude (3 m$\Omega$) approximately corresponds to the core orbital motion whose radius is about $0.3r \approx 200$ nm; this amplitude is in the same range as the result of the simulation shown in Fig. 2(c). The radius dependence of the resonance frequency is well reproduced by the simulation, as shown in Fig. 4(b). In particular, for the dots with $r$ = 700 nm, a fair agreement is observed. The systematic deviation between the experiments and simulation is possibly due to the inhomogeneous current distribution in the samples, which is more pronounced for the smaller dots.

Thus, we have demonstrated that a magnetic vortex core can be resonantly excited by an AC electric current. This phenomenon will facilitate detailed studies on the spin-transfer effect because of the simplicity of the system. The vortex core experiences a well-defined potential that is dependent on the dot shape; this potential is not sensitive to edge roughness. State-of-the-art time-resolved imaging techniques [19,20,25] can reveal the trajectory of the vortex core during excitation; this would lead to a better quantitative understanding.

The present work was partly supported by MEXT Grants-in-Aid for Scientific Research in Priority Areas and JSPS Grants-in-Aid for Scientific Research.

Figure Captions

Figure 1

Scanning electron microscope image of the sample along with a schematic configuration used for the measurements. The detection of the vortex excitation was performed by resistance measurements with a lock-in technique (223 Hz and current $I_{mes}$ = 15 µA) under the application of an AC excitation current $I_{exc}$ = 3 × $10^{11}$ A/m$^2$.

Figure 2

(a) Time evolution of the vortex under the AC current application. Magnetization direction $\boldsymbol{m}$ = ($m_x$, $m_y$, $m_z$) inside the dot on the $xy$ plane was obtained by micromagnetic simulation. The 3D plots indicate $m_z$ with the $m_x$ – $m_y$ vector plots superimposed. The plot on the left represents the initial state of the vortex core situated at the center of the dot with $r$ = 410 nm. The 3D plots on the right show the vortex on the steady orbital at $t$ = 80.6, 81.5, and 82.3 ns after applying the AC current ($f_0$ = 380 MHz and $J_0$ = 3 × $10^{11}$A/m$^2$). These plots are close-ups of the square region around the dot center indicated by the black square in the plot on the left. The time evolution of the core orbital from $t$ = 0 to 100 ns is superimposed only on the $t$ = 82.3 ns plot.

(b) Time evolutions of the vortex core displacement ($x$) for three excitation frequencies $f$ = 250, 340, and 380 MHz ($r$ = 410 nm and $J_0$ = 3 × $10^{11}$A/m$^2$).

(c) Radius of the steady orbital as a function of the frequency for the dots with $r$ = 410, 530, and 700 nm.

Figure 3

(a) Results of magnetoresistance measurements at room temperature for the dot with $r = 700$ nm. The red (blue) line is the result for $H \,/\!/\, I$ ($H \perp I$). The results are plotted as a deviation in resistance ($\Delta R_{/\!/}$ and $\Delta R_\perp$) from the state in the zero magnetic field where the core exists at the center of the dot. The spin structure in the dot for each state (at $\pm 150$ Oe denoted by the closed circles), which was determined by micromagnetic simulation, is also indicated.

(b) $|\Delta R_{/\!/}| - |\Delta R_\perp|$ as a function of the magnetic field.

Figure 4

(a) Experimental detection of the current-driven resonant excitation of a magnetic vortex core. The resistances are indicated as a function of the frequency of the AC excitation current for the dots with three different radii $r = 410$, 530, and 700 nm.

(b) Radius dependence of the resonance frequency. The blue rectangles and the red circles indicate the simulation and the experimental results, respectively. The experimental results for 8 samples are plotted. The red dashed line is the averaged value of the experimental data.

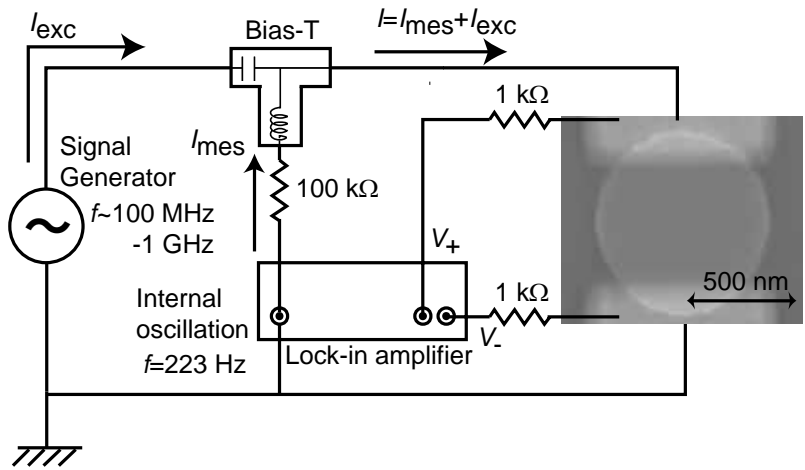

$I_{exc}$

Bias-T

$I = I_{mes} + I_{exc}$

1 kΩ

Signal
Generator
$f \sim 100$ MHz
-1 GHz

$I_{mes}$

100 kΩ

$V_+$

500 nm

Internal
oscillation
$f = 223$ Hz

$V_-$

Lock-in amplifier

1 kΩ

Fig. 1

Fig. 2

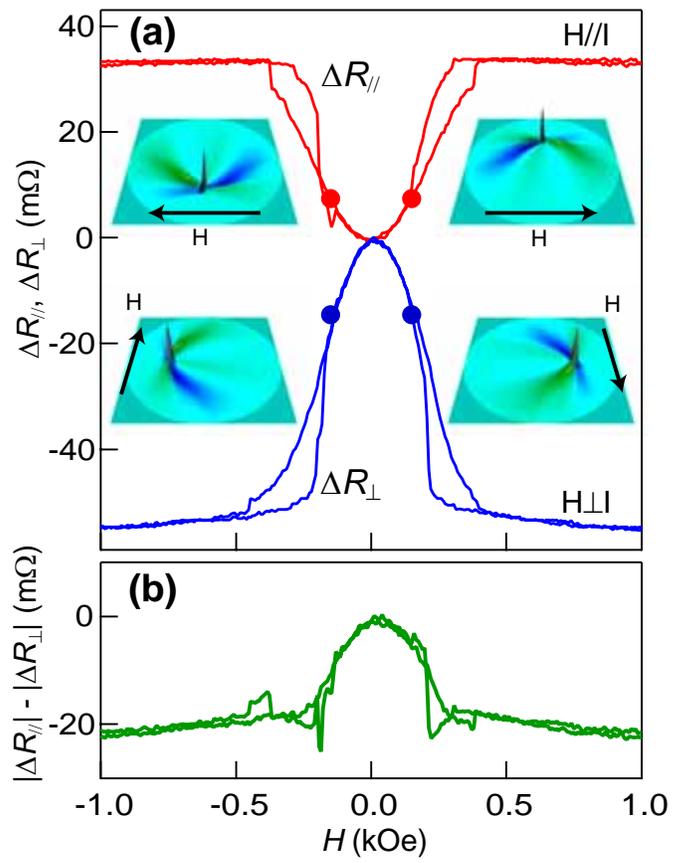

**(a)** $\Delta R_{//}$ $H//I$

$\Delta R_{//}, \Delta R_{\perp}$ (m$\Omega$)

$\Delta R_{\perp}$ $H \perp I$

**(b)**

$|\Delta R_{//}| - |\Delta R_{\perp}|$ (m$\Omega$)

$H$ (kOe)

Fig. 3

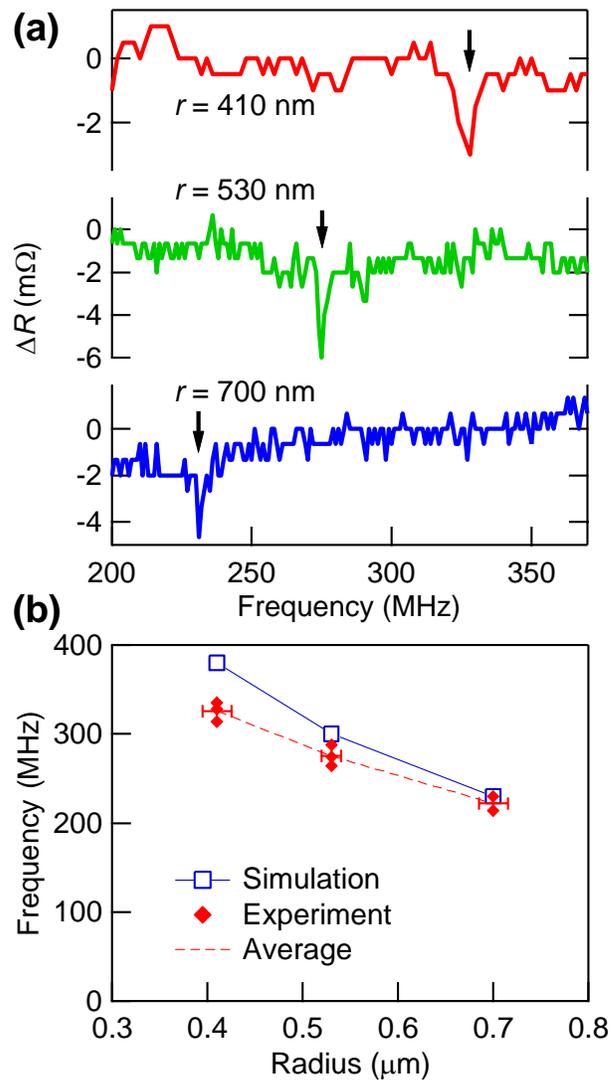

Fig. 4